\documentclass[12pt]{iopart}

\newcommand{\bml}{\numparts}
\newcommand{\eml}{\endnumparts}
\newcommand{\bey}{\begin{eqnarray}}
\newcommand{\eey}{\end{eqnarray}}
\newcommand{\be}{\begin{equation}}
\newcommand{\ee}{\end{equation}}
\newcommand{\dst}{\displaystyle}
\newcommand{\cA}{\mathcal{A}}
\renewcommand{\cal}{\mathcal}

\begin{document}


\title[Light-front gauge invariant formulation]{Light-front gauge invariant formulation and electromagnetic duality}

\author{J.A. Przeszowski}

\ead{jprzeszo@alpha.uwb.edu.pl} 

\address{Institute of Theoretical Physics, University of Bia{\l}ystok,\\
ul. Lipowa 41, 15-424 Bia{\l}ystok, Poland}

\date{\today}

\begin{abstract}
The gauge invariant formulation of Maxwell's equations and the
electromagnetic duality transformations are given in the light-front
(LF) variables. The novel formulation of the LF canonical
quantization, which is based on the kinematic translation generator
$P^{+}$ rather then on the Hamiltonian $P^{-}$, is proposed. 
This canonical quantization is applied for the free electromagnetic
fields and for the fields generated by electric and magnetic external
currents. The covariant form of photon propagators, which agrees with
Schwinger's source theory, is achieved when the direct interaction of
external currents is properly chosen. Applying the path integral
formalism, the equivalent LF Lagrangian density, which depends on two
Abelian gauge potentials, is proposed. 
Some remarks on the Dirac strings and LF non local structures are
presented in the Appendix.
\end{abstract}

\pacs{11.10.Ef, 11.15.Bt, 11.30.Er, 12.20.Ds}

\maketitle


\section{Introduction}

The LF formulation of Maxwell's electromagnetic theory with both the
electric and magnetic external currents is interesting at least for two
reasons: first as a relativistic field theoretical model and second as an
introduction to the dual description of QCD. The crucial step for such a
formulation is the electromagnetic duality transformation, which  allows
one to add magnetic currents into the electromagnetic system with
electric currents.
 
The first attempt in this direction, restricted to a classical
theory, done by Gambini and Salam\'o \cite{GambiniSalamo1979}, is partially
successful. They have selected two gauge invariant independent degrees of
freedom and they have formulated the continuous LF electromagnetic duality
transformation in terms of these degrees of freedom. They have shown that this
duality transformation is consistent with the usual duality transformation for 
electric and magnetic fields $\vec{E}$ and $\vec{B}$. However when trying to 
incorporate the Cabibbo-Ferrari theory \cite{CabibboFerrari1966} into the LF
formulation, they found incorrectly 4 gauge invariant independent degrees of
freedom. Also Gambini and Salam\'o have proposed effective Lagrangians with
the noncanonical current terms.

Next, some 5-6 years ago, there were two independent attempts to
describe the discrete electromagnetic duality within the LF description
of  Maxwell's theory. They  were initiated  by
Susskind \cite{Susskind96}, who has argued that the LF
electromagnetic duality transformation can be expressed simply in
terms of the transverse potentials as 
\be
A_i \longmapsto - \epsilon_{ij}A_j.\label{Eq1}
\ee
This conjecture has been explicitly checked by Brisudova \cite{Brisudova99},
within the LF canonical quantization in the LC gauge $A_{-} = A^{+}
= 0$, with a final conclusion that only for free fields one can
define such duality. This follows directly from the
starting point, since any one gauge potential description of electromagnetic
theory is evidently false for a theory with
electric and magnetic currents. Instead one should use either a two
gauge potential approach or a LF version of Dirac strings.
Also the idea of entanglement of duality with gauge transformations
\cite{Susskind96},  \cite{Brisudova99} is physically misleading - these
transformations are fundamentally different and should be treated separately. 

The next attempt, by Mukherjee and Bhattacharya \cite{Asmita2000}, is
based on a different approach with two gauge potentials. This
method, introduced long ago by Zwanzinger \cite{Zwanziger1971} within
the equal-time (ET) formulation, leads to a quite complicated picture,
where one has to introduce a constant space-like vector $n^\mu,
n^2 < 0$, which breaks the Lorentz invariance of starting Lagrangian
density. This Lagrangian leads to many constraints, which follow from 
two gauge symmetries. Mukherjee and Bhattacharya, implementing the Dirac canonical
quantization procedure for systems with constraints \cite{Dirac1958}, found 
that effectively only one component of each
gauge potential is an independent degree of freedom. Therefore after
redefining these independent modes as two transverse components of
some effective gauge potential they prove the Susskind conjecture
(\ref{Eq1}) for the interacting electromagnetic theory with both
electric and magnetic external currents. Their analysis of the quantum
theory ends with the structure of LF commutators and Hamiltonian.
Since their Hamiltonian contains a rotationally noninvariant term, which
describes instantaneous interaction of electric and magnetic currents, then
we may worry if a covariant perturbation theory follows from their analysis.

The aim of our paper is to formulate a clear LF description
of  Maxwell's theory, where only the gauge invariant objects are
used. We will start with the free LF Maxwell equations and define
the LF electromagnetic duality transformation as their symmetry.
Then we will quantize the system canonically, treating $x^{+}$ as
the LF evolution parameter and using $\bar{x} = (x^{-}, x^2, x^3)$
as coordinates on the LF (hyper-)surface. The Poincare generators will be 
defined in terms of the symmetric energy-momentum tensor $T^{\mu
\nu}_{sym}$, which is gauge invariant, rather than the canonical
energy-momentum tensor $T^{\mu \nu}_{can}$, which is a gauge
dependent object. We will propose a novel canonical procedure for the
LF systems, where the generator of translations in $x^{-}$ direction is
used for deriving canonical LF Poisson brackets for all independent
fields.  Since this generator is a kinematic one, it will keep its 
free field form also for an interacting theory, where a Hamiltonian
contains some interaction part. Our canonical procedure is
explicitly duality invariant and we will prove the Susskind conjecture 
(\ref{Eq1}), but for fields with a different physical interpretation.

Our paper is organized as following. 
In section 2 we present a very concise ET description of Maxwell's
equations with external electric and magnetic sources.
In section 3 we start with the LF tensor formulation  of Maxwell's equations and the electromagnetic duality transformation. 
We also introduce the LF notation for the electromagnetic fields.
In section 4 we analyze the free field case, when
no external sources are present. In section 5 we consider the
general case of the Maxwell equations, when both kinds of external
sources are present, paying a special attention to the LF duality
transformation. In section 6 we switch our analysis to the path integral
formulation and derive all propagators, which mediate interactions between electric
and magnetic currents. We also find a local Lagrangian, which is equivalent to
the proposed Hamiltonian formulation. In conclusions we discuss our results
indicating the crucial points and suggesting possible further investigations. The LF
notation and the Green functions are presented in the Appendix A. In the Appendix B
we give a short presentation of different Dirac's strings for the ET and LF
formulations. 

\section{Vector notation for Maxwell's equations}

Before starting our novel LF approach, let us briefly review
the ET formulation. The magnetic charges
$\rho_{m}$ and currents $\vec{J}_m$ are formally added to the
electric charges $\rho_e$ and currents $\vec{J}_e$, when one writes
Maxwell's equations 
\bml \label{conMaxeq}
\bey
\vec{\nabla} \times \vec{B} & = \frac{\partial}{\partial t}
\vec{E} + \vec{J}_e,   \quad     \qquad 
\vec{\nabla} \cdot \vec{E}  & = \rho_{e},\label{Eq2a}\\
\vec{\nabla} \times \vec{E} & =  - \frac{\partial}{\partial t} \vec{B} -
\vec{J}_m,                  \qquad 
\vec{\nabla} \cdot \vec{B}  & = \rho_m.\label{Eq2b}
\eey \eml
This set of equations is invariant under the electromagnetic
duality transformation 
\bml
\bey
\vec{E} & \longmapsto & \vec{B} \longmapsto - \vec{E},
\label{ETdt}\\
\vec{J}_e & \longmapsto & \vec{J}_m \longmapsto - \vec{J}_e,\label{Eq3b}\\
\rho_e & \longmapsto & \rho_m \longmapsto - \rho_e.\label{Eq3c}
\eey
\eml 
This vector notation is not a suitable starting point for the LF
formulation, since for 3-vectors, contrary to 4-vectors, there is no
definition of LF components. 

Usually, when there are no magnetic sources $(\rho_{m} = \vec{J}_{m} = 0)$,  then a pair of homogeneous Maxwell's equations (\ref{Eq2b}), may
be removed with the help of 4-vectors $A_\mu = (A_0, \vec{A})$ by the standard definitions of gauge field potentials: 
\be
\vec{E} = - \vec{\nabla}A_{0} - \frac{\partial}{\partial t}
\vec{A}, \qquad \vec{B} = \vec{\nabla} \times \vec{A}.
\ee 
This allows to express the inhomogeneous Maxwell equations (\ref{Eq2a})
in terms of 4-vectors $A_\mu$, which further can be easily converted
into the LF notation. This may explain why, in almost all papers on
the LF quantization of  Maxwell's theory, only the gauge
field potential approach has been used (for reviews see
\cite{BrodskyPauliPinsky1993}, \cite{BrodskyPauliPinsky1997}).

However if magnetic sources are present, then the
formulation in terms of a single gauge field potential runs into
inconsistencies. The way out is either by means of the Dirac string
\cite{Dirac1931}, \cite{Dirac1948} or by the Wu-Yang potentials
\cite{WuYang1975}, to mention only the best known solutions.
But we would like to stress that the gauge potentials are only quite
useful, but are not necessary, for a consistent quantization
of Maxwell's equations (\ref{Eq2a}, \ref{Eq2b}) within the ET approach
\cite{Maxwellquant1}, \cite{Maxwellquant2}. Actually,  they
couple locally to external currents, but the canonical
commutation relations can be solely expressed in terms of the
electromagnetic fields $(\vec{E}, \vec{B})$. Thus we expect that
also within the LF approach one can consistently quantize 
Maxwell's equations directly in terms of electromagnetic fields.


\section{LF notation for Maxwell's equations and electromagnetic duality}


As a starting point for the LF formulation of Maxwell's
electromagnetism with classical external electric sources, we take the
tensor formulation:
$ \partial_{\mu} F^{\mu \nu} = J^{\mu},\ \epsilon^{\mu \nu \lambda \rho} \partial_{\nu} F_{\lambda \rho} = 0$.
These equations can be easily transformed into the LF coordinates as
the inhomogeneous LF Maxwell equations:
\bml \label{LFinhomMaxwell}
\bey 
\partial_{+} E_{-} & = & \partial_{i} E_{i} + J^{-},\label{LFinhomEi}\\
\partial_{+} B_i & = & - \partial_{-} E_i - \epsilon_{ij}
\partial_j B_{-} + J^i, \label{LFinhomBi}\\
\quad 0 & = & \partial_{-} E_{-} + \partial_i B_i + J^{+}, \label{LFelgauss}
\eey 
\eml
and the LF Bianchi identities: 
\bml \label{LFhomMaxwell}
\bey
\partial_{+} B_i & = & \partial_{-} E_i - \partial_i E_{-},\label{LFhomEi}\\
\partial_{+} B_{-} & = & \epsilon_{ij} \partial_{i} E_{j}, \label{LFhomBi}\\
\partial_{-} B_{-} & = & \epsilon_{ij} \partial_{i} B_{j}, \label{LFmaggauss}
\eey \eml
where the LF electromagnetic fields are defined as
$E_{-} = F_{+-},\ E_i = F_{+i},\ B_{i} = F_{-i},\ \epsilon_{ij} B_{-} = F_{ij}.$ These equations, being manifestly gauge invariant, may be taken as the basis for the further canonical procedure, both for the free field case and the interacting field case. \\
Next, the tensor form of the electromagnetic duality transformations 
\be
F_{\mu \nu} \longmapsto \star F_{\mu \nu} = \frac 1 2 \epsilon_{\mu
\nu  \lambda \sigma} F^{\lambda \sigma},
\ee
can be rewritten in the LF coordinates as 
\bml \label{LFdualitytr}
\bey
E_{-} & \longmapsto  B_{-},               \qquad B_{-} &\longmapsto  -   E_{-}, \label{eq:8a}\\
E_{i} & \longmapsto  \epsilon_{ij} E_{j}, \qquad  B_{i} & \longmapsto - \epsilon_{ij} B_{j},\label{strangeduality}
\eey
\eml
and hereafter we will refer to them as the LF electromagnetic duality transformation. If one doubts that (\ref{strangeduality}) is correct, then one may introduce another notation:
\bml
\be
\bar{E}_2 = E_2,\quad  \bar{E}_3 = - B_2, \quad \bar{B}_2 = E_3, \quad \bar{B}_3 =  B_3,
\ee
which allows to express (\ref{strangeduality}) as 
\be
\bar{E}_{i} \longmapsto  \bar{B}_{i},  \qquad \bar{B}_{i} \longmapsto  -   \bar{E}_{i}.
\ee
\eml
However we think that the very form of the LF electromagnetic duality
is not important - what really matters is how these transformations
act on Maxwell's equations. One can easily convince himself, that in
the absence of external sources $(J^{\mu}=0)$ the LF Maxwell equations
(\ref{LFinhomEi}, \ref{LFinhomBi}, \ref{LFelgauss}) and (\ref{LFhomEi}, \ref{LFhomBi}, \ref{LFmaggauss}) transform, mutually in pairs, under 
(\ref{eq:8a}, \ref{strangeduality}). Since the duality transformation
for sources (\ref{Eq3b}, \ref{Eq3c}) can be directly expressed in terms
of the 4-currents $J^{\mu} = (\rho_e, \vec{J}_e)$ and $K^{\mu} = (\rho_m,\vec{J}_m)$ 
\bml
\be \label{Eq4}
J^{\mu} \longmapsto K^{\mu} \longmapsto - J^{\mu}, 
\ee
then for the LF components one has 
\be%
J^{\pm}  \longmapsto  K^{\pm}  \longmapsto  - J^{\pm}, 
\qquad J^{i} \longmapsto  K^{i} \longmapsto  - J^{i}.\label{eq:8b}
\ee
\eml
At last, we also need a gauge invariant energy-momentum tensor, thus 
we take the symmetric energy-momentum tensor
\be
T_{sym}^{\mu \nu} = F^{\mu}_{\ \lambda} F^{\lambda \nu} + 
\frac 1 4 g^{\mu \nu} F^{\lambda \rho} F_{\lambda \rho},
\ee
with the LF components: 
\bml \label{symEMtensor}
\bey
\fl T_{sym}^{+-} & = & \frac 1 2 \left( E_{-}^2 + B_{-}^2 \right),\hspace{15pt}\
T_{sym}^{++} = B_i^2, \hspace{15pt} \ T_{sym}^{+i} =E_{-} B_i -
\epsilon_{ij}B_j B_{-},\\ 
\fl  T_{sym}^{-i}& =& - E_{-} E_i -
\epsilon_{ij} E_j B_{-},\  \hspace{5pt} \ \ T_{sym}^{--}  =  E_i^2, \\
\fl T_{sym}^{ij} & = &  - \left( E_i B_j + E_j B_i\right) + \delta_{ij}
\left( E_k B_k + \frac 1 2 E_{-}^2 + \frac 1 2 B_{-}^2 \right).
\eey
\eml 
These components of $T_{sym}^{\mu \nu}$ are invariant under the LF
duality transformations (\ref{eq:8a}, \ref{strangeduality}) and in our
further investigations we will always keep our quantization procedure
both the gauge and duality invariant.


\section{Gauge invariant canonical quantization for free fields}

In this section we would like to focus our attention on the quantization of
electromagnetic degrees of freedom, thus we will consider the case of free
electromagnetic fields when all external sources vanish ($J^\mu = 0$). 
In our LF canonical procedure we choose $x^{+}$ as the temporal
evolution parameter, thus all classical LF Maxwell equations
(\ref{LFinhomEi} - \ref{LFmaggauss}) should be classified either as
equations of motion or constraints. Usually, when an equation contains
a term with the temporal partial derivative $\partial_{+}$ it is 
classified as an equation of motion, otherwise it is a constraint. 
However here situation is more tricky, since both equations
(\ref{LFinhomBi}, \ref{LFhomEi}) contain terms $\partial_{+}B_i$,
thus their linear combination give rise to the effective
constraints 
\be
2 \partial_{-} E_i =  \partial_i E_{-} -  \epsilon_{ij}
\partial_j B_{-}, \label{eqforEi}
\ee
and the effective equations of motion 
\be
2 \partial_{+} B_i  =  - \partial_i E_{-} - \epsilon_{ij} \partial_j B_{-}.
\ee
Thus we conclude, that for Maxwell's theory there are more LF
constraints then ET ones. A formal reason for this difference follows 
from the inhomogeneous equation (\ref{LFinhomBi}), which being an Euler-Lagrange
equation, generally contains the temporal derivative term $\partial_{+} D_i$, where 
the canonical momentum field is defined as 
\bml
\be
D_i = \frac{\partial {\cal L}}{\partial E_i}.
\ee   
On contrary, the homogeneous equation (\ref{LFhomBi}), being a Bianchi
identity, contains the term $\partial_{+} B_i$. Generally, these two fields
($D_i, B_i$) form a canonical pair and are independent degrees of
freedom. However, for Maxwell's theory, with the Lagrangian density
\be
{\cal L} = - \frac 1 4 F_{\mu \nu} F^{\mu \nu} = \frac 1 2 \left(E_{-}^2 - B_{-}^2\right) + E_i B_i,
\ee
one finds that these the canonical variables $(D_i, B_i)$ are
constrained 
\be
 D_i = B_i, \label{canconstr}
\ee
\eml 
and two different equations contain the same term
$\partial_{+} B_i$. Further consequences of this observation will
be presented elsewhere and here we only stress that the constraint
(\ref{canconstr}) is gauge invariant, thus truely a physical phenomenon, 
which can be intimately connected with the Lorentz symmetry of Maxwell's
electrodynamics.
  
Since there is no equation of motion for $E_i$, then we would like to 
treat these fields as dependent field variables and remove them, from
our canonical analysis, by means of the constraint equations
(\ref{eqforEi})\footnote{This is quite similar to the nondynamical components of fermion field $\psi_{-}, \psi^{\dag}_{-}$, which are removed by solving the constraint part of Dirac's equations.}.

All remaining LF electromagnetic fields $(E_{-}, B_{-}, B_i)$
have their  equations of motion
\bml 
\bey
\left(2 \partial_{+} \partial_{-} - \triangle_{\perp}\right)E_{-} &
= & 0,\label{effeqEmfree}\\
\left(2 \partial_{+} \partial_{-} - \triangle_{\perp}\right)B_{-} &
= & 0,\label{effeqBmfree}\\
\qquad \qquad 2 \partial_{+} B_i & = & - \partial_i E_{-} - \epsilon_{ij}
\partial_{j} B_{-},\label{effeqBifree}
\eey
\eml
and also appear in the constraints\footnote{This is quite analogous to the status of ET electromagnetic fields $(\vec{E},\vec{B})$.}
\bml \label{effcontsrfree}
\bey
\partial_{-} E_{-} + \partial_i B_i & = & 0,\label{constrJpfree}\\
\partial_{-} B_{-} -  \epsilon_{ij} \partial_{i} B_{j} & = & 0.\label{eq:14b}
\eey
\eml 
Further we  notice, that only these LF fields appear in the gauge invariant
Poincare generators of translations
\bml
\bey
{P}^{-} & = & \int d^3\bar{x} \ {T}^{+-}_{sym} = \int d^3\bar{x} \ \ 
\frac 1 2 \left({E}_{-}^2 + {B}_{-}^2 
\right),\\
{P}^{+} & = & \int d^3\bar{x} \ {T}^{++}_{sym} = \int
d^3\bar{x} \ {B}_i^2 ,\\ 
P^{i} & = & \int d^3\bar{x} \ T^{+i}_{sym}
= \int d^3\bar{x} \ \left (E_{-} B_i - 
\epsilon_{ij}B_j B_{-}\right).
\eey
\eml 
Therefore, using the equations of motion (\ref{effeqEmfree}, \ref{effeqBmfree},
\ref{effeqBifree}), one can prove that these generators are the constants of motion
(do not depend on $x^{+}$). 
 
Since our system contains constraints for canonical field variables,
then our next steps within the canonical quantization procedure could be
taken according to either the Dirac method \cite{Dirac1958} or the
Faddeev-Jackiw (FJ) method of the Hamiltonian reduction
\cite{FaddeevJackiw}. However, we prefer to choose a less technical procedure
and solve the constraint equations (\ref{constrJpfree}, \ref{eq:14b}) by a suitable 
parameterization of electromagnetic fields:
\be
{E}_{-}  =  - \partial_i {\cal A}_i,\qquad {B}_{-}  =  \epsilon_{ij} \partial_i {\cal A}_j,\qquad {B}_i = \partial_{-} {\cal A}_i, \label{LFparameterazation}
\ee
where  ${\cal A}_i$ are the independent gauge invariant fields. This
parameterization clearly shows, that within the LF formulation, there are only 2
independent dynamical modes with the equations of motion 
\be
\left(2 \partial_{+} \partial_{-} - \triangle_{\perp}\right){\cal A}_{i} = 0  \label{eqAifree}.
\ee
Thus we have a complete agreement between the ET and LF formulations: in both
approaches we have 2 independent relativistic modes, which are described by 4 canonical ET fields and 2 canonical LF fields.
 
This observation agrees with \cite{GambiniSalamo1979}, though Gambini and 
Salam\'o take $E_{-}$ and $B_{-}$ (using a different notation for them) as the
independent electromagnetic modes. Also it forms the physical explanation of the
additional constraints (\ref{eqforEi}), which appear in the LF approach - there is
only 1 independent canonical field variable for any relativistic independent mode,
contrary to the ET approach, where every mode is described by 2 canonical fields.

Using our parameterization of fields, we may express the gauge invariant 
Poincare generators as 
\bml \label{freeAmomep}
\bey
{P}^{-} & = & \int d^3\bar{x} \
{T}^{+-}_{sym} = \frac 1 2 \int d^3\bar{x} \ \left[ \left(\partial_i {\cal
A}_{i}\right)^2 + \left( \epsilon_{ij} \partial_i {\cal
A}_{j}\right)^2 \right], \label{P-free}\\
{P}^{+} & = & \int d^3\bar{x} \ {T}^{++}_{sym} = \int
d^3\bar{x} \ \left( \partial_{-} {\cal A}_i\right)^2 ,\label{P+free}\\ 
P^{i} & = & \int d^3\bar{x} \ T^{+i}_{sym}
= - \int d^3\bar{x} \ \left (\partial_i {\cal A}_{j} 
\partial_{-}{\cal A}_j\right),
\eey
\eml 
while the LF duality electromagnetic transformations (\ref{eq:8a}, \ref{strangeduality}) boil down to
\be
{\cal A}_i \mapsto - \epsilon_{ij} {\cal A}_j, \label{dualitysqAi}
\ee 
which agrees with Susskind's conjecture (\ref{Eq1}). 
Here we notice that our results have the same form as those found 
within the usual gauge potential approach for the the LC gauge
condition $A_{-} = A^{+} = 0$, provided we set ${\cal A}_i = A_i$. We
stress that it is just a coincidence, since our variables ${\cal A}_i$, being gauge invariant
quantities, are true physical fields and there is no gauge
transformation for them. For free Maxwell's theory, when one chooses
the LC gauge condition $A_{-} = A^{+} = 0$, all nonphysical modes are
removed and the transverse the potentials $A_i$ describe physical
modes. However, for an interacting theory, ignoring the difference 
between ${\cal A}_i$ and $A_i$, one may run into serious difficulties
- just like in \cite{Brisudova99}.

Also we would like to indicate another difference between the gauge
independent field variables ${\cal A}_i$ and the usual gauge field
potential $A_\mu$. In our case, we use ${\cal A}_i$ to solve
identically two LF constraints (\ref{constrJpfree}, \ref{eq:14b}),
while the usual vector gauge potentials $A_\mu$ solve identically all
Bianchi identities (\ref{LFhomMaxwell}, \ref{LFhomEi}, \ref{LFhomBi},
\ref{LFmaggauss}). 

Since we have effectively reduced the constrained system $(E_{-}, B_{-}, B_{i})$
into the independent dynamical fields ${\cal A}_i$, then the LF canonical 
procedure may follow directly from the Poisson bracket relation
\be
\partial_{+} {\cal A}_{i} = \left\{ {\cal A}_{i}, P^{-}\right\}_{PB},
\ee
when we take the equations of motion (\ref{eqAifree}) and the LF 
Hamiltonian $P^{-}$ (\ref{P-free}). 
However here we would like to propose a novel LF canonical procedure, which uses
$P^{+}$ instead of $P^{-}$. Thus we argue that one may start with 
the trivial relation 
\be \label{kinpoissrel}
\partial_{-} {\cal A}_{i} = \left\{ {\cal A}_{i}, P^{+}\right\}_{PB},
\ee
and using the expression (\ref{P+free}) for $P^{+}$, one may easily infer the 
canonical LF Poisson bracket
\be
2 \left\{ \partial_{-} {\cal A}_{i}(x^{+}, \bar{x}), {\cal A}_{j}(x^{+},
\bar{y})\right\}_{PB} =   - \delta_{ij} \delta^{3}(\bar{x} - \bar{y}).\label{AiAifreebracket}
\ee
Since $P^{+}$ is the kinematic generator, then the relation
(\ref{kinpoissrel}) can be effectively used for finding the LF
canonical brackets also  for an interacting theory.

For the consistency check, one may use the LF canonical brackets
(\ref{AiAifreebracket}) to calculate other Poisson brackets 
\be
\partial^\mu {\cal A}_{i} = \left\{ {\cal A}_{i}, P^\mu\right\}_{PB},
\ee
finding (for $\mu = -$) the proper form of the equations of motion (\ref{eqAifree})
and (for $\mu = j$) the trivial identities.

The canonical quantization procedure means that canonical
variables are changed into quantum operators and Poisson brackets
are transformed into commutators\footnote{We will denote the quantum field
operators by the same symbols as the respective classical fields
hoping that this will not lead into any misunderstanding.} 
\be
2 \left[ \partial_{-} {\cal A}_{i}(x^{+}, \bar{x}), {\cal
A}_{j}(x^{+}, \bar{y})\right] =  - i \delta_{ij}
\delta^{3}(\bar{x} - \bar{y})\label{AiAjfreecommut}.
\ee
We stress that both the classical brackets and the quantum commutators
are invariant under the LF duality transformation (\ref{dualitysqAi}),
thus our canonical quantization procedure is both the gauge and duality invariant.


\section{Electric and magnetic external currents}

As a next step we would like to consider the case of LF
electromagnetic fields interacting with electric and magnetic currents.
Therefore we take the inhomogeneous Maxwell equations with electric
external currents (\ref{LFinhomEi}, \ref{LFinhomBi}, \ref{LFelgauss})
and applying the  duality transformations  (\ref{eq:8a}, \ref{strangeduality}, \ref{eq:8b}) we generate the equations with magnetic currents
\bml \label{genlfMaxeq}
\bey
\fl \partial_{+} E_{-} & = \partial_{i} E_{i} + J^{-},
\qquad \qquad \qquad \partial_{+} B_{-} & = \epsilon_{ij} \partial_{i} E_{j} + K^{-}, \label{eq:23a}\\ 
\fl \partial_{+} B_i & =  - \partial_{-} E_i - \epsilon_{ij}
\partial_j B_{-} + J^i, \qquad \partial_{+} B_i &  =
\partial_{-} E_i - \partial_i E_{-} 
+ \epsilon_{ij} K^j,\label{eqBi2}\\
\fl \partial_{-} E_{-} & = - \partial_i B_i  - J^{+}, \qquad 
\qquad \qquad \partial_{-} B_{-} & = \epsilon_{ij} \partial_i B_j  - K^{+}.\label{eq:23c}
\eey 
\eml
These LF Maxwell's equations with electric and magnetic currents are equivalent 
to (\ref{Eq2a}) and (\ref{Eq2b}) and they form the starting point for the 
analysis of interacting theory. There are two consistency conditions for these
Maxwell's equations: the electric current continuity equation 
\bml
\be \label{electconteq}
\partial_{+} J^{+}+ \partial_{-} J^{-} + \partial_{i} J^{i} = 0,
\ee
and the magnetic current continuity equation 
\be \label{magnetconteq}
\partial_{+} K^{+}+ \partial_{-} K^{-} + \partial_{i} K^{i} = 0.
\ee
\eml
In this paper, we will not analyze the dynamical structure of these electric 
and magnetic currents, thus we will only suppose that  these two conservation
laws always hold. We stress that all equations (\ref{eq:23a}, \ref{eqBi2},
\ref{eq:23a}) are inhomogeneous, thus we argue that generally no equation can
be interpreted as a Bianchi identity and there is no one gauge potential 
approach, which would lead to these equations.

We will carry our LF quantization procedure following the same steps as in the
previous section and start with the identification of equations of motion and
constraints. As before $E_i$ is a nondynamical field variable, which now can 
be determined from the effective constraint
\be
2 \partial_{-} E_i = \partial_i E_{-} -  \epsilon_{ij}
\partial_j B_{-} + J^i - \epsilon_{ij} K^j.\label{eqEi}
\ee
The other fields are dynamical with the effective equations of motion 
\bml
\bey
\qquad \qquad 2 \partial_{+} B_i & = & - \epsilon_{ij} \partial_j B_{-} -
\partial_{i} E_{-} + J^{i} + \epsilon_{ij} K^j, \label{moneqbi}
\\
\left( 2\partial_{+} \partial_{-} - \triangle_{\perp} \right) B_{-}
& = & \epsilon_{ij}\partial_{i} J^{j} + \partial_i K^i + 2 \partial_{-}
K^{-},\\ 
\left( 2\partial_{+} \partial_{-} - \triangle_{\perp} \right) E_{-} & = &  -
\epsilon_{ij}\partial_{i} K^{j} + \partial_i J^i + 2 \partial_{-}
J^{-}, \label{moneqem}
\eey
\eml 
while the remaining constraint equations are following
\bml \label{elmagconstr}
\bey
\partial_{-} E_{-} + \partial_i B_i & = & - J^{+},\label{eq:27a}\\
\partial_{-} B_{-} - \epsilon_{ij} \partial_i B_j & = & - K^{+}.\label{eq:27b}
\eey
\eml
These last equations can be interpreted as the LF electric Gauss law
and the LF magnetic Gauss law, respectively. Similarly, to the previous section, 
we wish to find a useful parameterization of the LF electromagnetic fields,
which would solve identically the above constraints. However, now the best choice is no longer evident, since we have to allow for some explicit dependence of electromagnetic
fields on the external LF charges $J^{+}$ and $K^{+}$. Therefore we may try
different possibilities, for example, we may take either
\bml
\bey
E_{-} & = & - \partial_i {\cal A}_i - (\partial_{-})^{-1} J^{+},\label{eq:28a} \\
B_{-} & = & \epsilon_{ij} \partial_i {\cal A}_j - (\partial_{-})^{-1} K^{+},\label{eq:28b}\\
B_{i} & = & \partial_{-} {\cal A}_i, \label{eq:28c}
\eey
\eml
or 
\bml
\bey
E_{-} & = & - \partial_i {\cal A}_i,\label{eq:29a} \\
B_{-} & = & \epsilon_{ij} \partial_i {\cal A}_j,\label{eq:29b}\\
B_{i} & = & \partial_{-} {\cal A}_i - \partial_i \Delta_{\perp}^{-1} J^{+} + \epsilon_{ij} \partial_j \Delta_{\perp}^{-1} K^{+}. \label{eq:29c}
\eey
\eml
Apparently this arbitrariness is annoying as an ambiguity of our
approach, however we interpret it as a manifestation of different
possible choices of independent modes for the interacting system.
Every different choice is self-consistent and various choices should
ultimately lead to the same physical predictions, though they may
differ at the intermediate stages. Our novel canonical procedure
offers some extra hints for a choice of parameterization. If we demand
that the translation generator $P^{+}$ remains its free field forms 
\be
{P}^{+} =  \int d^3\bar{x} \ {B}_i^2 = \int
d^3\bar{x} \ \left( \partial_{-} {\cal A}_i\right)^2, \label{intPplus}
\ee
then the only acceptable parameterization for $B_i$ is (\ref{eq:28c}).
The nonlocal (in $x^{-}$) dependence of electromagnetic fields 
$E_{-}$ and $B_{-}$ on $J^{+}$ and $K^{+}$, given by (\ref{eq:28a})
and (\ref{eq:28b}), can be interpreted as constant vector strings.
Since these strings appear while we solve the Gauss law constraint
equations, we may call them the Gauss law strings. A more
detailed discussion of these strings and their relation to the usual Dirac strings is presented in the \ref{appstrings}.
 
Now it is quite an easy exercise to check that all equations of motion (\ref{moneqbi} - \ref{moneqem}) are equivalent to the equation of motion for ${\cal A}_i$ fields
\bey
\left( 2 \partial_{+} \partial_{-} - \Delta_{\perp} \right){\cal A}_{i} & = & \epsilon_{ij} \partial_{j} (\partial_{-})^{-1} K^{+}  + \partial_{i}
(\partial_{-})^{-1} J^{+}  + J^i + \epsilon_{ij} K^j,   \label{eq:35}
\eey
provided the covariant conservations laws for the electric and 
magnetic currents (\ref{electconteq}, \ref{magnetconteq}) are taken
into account. Thus we see that in our gauge invariant description of
Maxwell's theory with electric and magnetic external currents there
are still 2 independent dynamical degrees of freedom, just like in the
free field case, though now we have also constant string structure
$(\partial_{-})^{-1}$ for the interaction terms. 

The canonical Poisson brackets for ${\cal A}_i$ fields follow immediately from the trivial equation
\be
\partial_{-} {\cal A}_i = \left\{ {\cal A}_i, P^{+} \right\}_{PB},
\label{quantcond}
\ee
with $P^{+}$ given by (\ref{intPplus}), leading to 
\be \label{AiAiintbracket}
2 \left\{ \partial_{-} {\cal A}_{i}(x^{+}, \bar{x}), {\cal A}_{j}(x^{+},
\bar{y})\right\}_{PB} =   - \delta_{ij} \delta^{3}(\bar{x} - \bar{y}),
\ee
which is precisely the free field result. Similarly, also $P^i$ 
is a kinematic generator 
\be
P^{i} =  - \int d^3\bar{x} \ \left (\partial_i {\cal A}_{j}
\partial_{-}{\cal A}_j\right),
\ee
but $P^{-}$, being the LF Hamiltonian, should contain interaction
terms. The very form of $P^{-}$ should be consistent with the 
effective equation of motion (\ref{eq:35}), which we write as 
the Poisson bracket equation
\bml
\be
\fl 2 \partial_{+} \partial_{-} {\cal A}_i = 2 \left\{ \partial_{-} {\cal A}_i, P^{-} \right\}_{PB} =  \Delta_{\perp} {\cal A}_{i} +  \epsilon_{ij} \partial_{j} (\partial_{-})^{-1} K^{+}  + \partial_{i}
(\partial_{-})^{-1} J^{+}  + J^i + \epsilon_{ij} K^j.
\ee
Next, due to (\ref{AiAiintbracket}), we infer 
the functional differential equation for $P^{-}$
\be
- \frac{\delta P^{-}}{\delta {\cal A}_i} =  \Delta_{\perp} {\cal A}_{i} +  \epsilon_{ij} \partial_{j} (\partial_{-})^{-1} K^{+}  + \partial_{i}
(\partial_{-})^{-1} J^{+}  + J^i + \epsilon_{ij} K^j,
\ee
\eml
which can be solved as  
\bey
{P}^{-} & = & \frac 1 2 \int d^3\bar{x} \left[ (\partial_i \cA_i)^2 + (
\epsilon_{ij} \partial_i \cA_j )^2 \right] \nonumber \\
&& 
+ \int d^3\bar{x} \left[ \partial_i \cA_i (\partial_{-})^{-1} J^{+} 
- \epsilon_{ij} \partial_i \cA_j (\partial_{-})^{-1} K^{+}\right] \nonumber\\
&& - \int d^3\bar{x} \cA_i (J^i + \epsilon_{ij} K^{j})  + H_{cur}.
\label{totalHamelmag}
\eey
$H_{cur}$ is a constant of functional integration, which may depend on
external currents $J^\mu$ and $K^\mu$ and we will call it the current
Hamiltonian.

We notice that the above canonical procedure is invariant under the LF
electromagnetic duality transformation, which again has the simple form 
for the independent modes  
\be
{\cal A}_{i}  \longmapsto  - \epsilon_{ij} {\cal A}_{j}, \label{gaugeinvduality}
\ee
provided the current Hamiltonian in (\ref{totalHamelmag}) is also
invariant under the duality transformation of external
currents (\ref{eq:8b}). Actually, the determination $H_{cur}$ is not a
trivial task and lies beyond the scope of the canonical quantization
procedure. In the next section, we will demand a covariant form of the
perturbative propagators and this will uniquely fix $H_{cur}$. 

As a concluding remark of this section, we stress that even though the
expression for the LF Hamiltonian (\ref{totalHamelmag}) is quite
similar to the well known LF Hamiltonian for the LC gauge $A_{-} = 0$,
then this similarity is quite misleading - one must be aware that
there is no way to derive this Hamiltonian within the one gauge
potential approach and this explains the failure of \cite{Brisudova99}
for the case of interacting Maxwell's theory.


\section{Path integral for electric and magnetic sources}


The canonical structure, which we have found in the previous section,
forms a good starting point for the canonical quantization procedure
for the sector of electromagnetic fields. There is yet an undetermined
current Hamiltonian $H_{cur}$, which describes the LF instantaneous
interaction of currents and we have to fix it ultimately. We believe
that the simplest way to find the consistent current Hamiltonian
$H_{cur}$ follows from the structure of perturbative propagators. 
This is based on the observation, that within the LF formulation, 
quite frequently one has to supplement a chronological (in $x^{+}$)
product of quantum field operators by some LF instantaneous
contribution from a Hamiltonian, when one finds a LF perturbative
propagator. Since perturbative propagators are derived most easily
from a path integral definition of a generating functional, 
therefore in this section we will depart from the canonical
quantization procedure but rather concentrate on the path integral
formulation.

We define the generating functional of all Green functions as the
phase-space path-integral, which follows naturally  from  the
canonical structure given in the previous section\footnote{We will
omit the normalization constants for all path-integrals which will
appear in this section keeping in mind the normalization condition
$Z[0] = 1$.} 
\bey \label{magpathint1}
\fl Z[J^{+}, J^{i}, K^{+}, K^{i}] & = & \int {\cal D} {\cal A}_i \
exp{\left(\ i \int d^4x \ \partial_{+} {\cal A}_i \partial_{-} {\cal A}_i\right)}\exp{\left(\ - i \int dx^{+} P^{-}\right)}. 
\eey 
Here we stress that the path integral field variables ${\cal A}_i$ 
are the gauge invariant quantities and they are the canonical 
variables in the unconstrained Hamiltonian phase-space.  Since all
integrals are Gaussian we may immediately perform them and 
get the result
\bey
Z[J^{+}, J^{i}, K^{+}, K^{i}] & = & \exp{\frac i 2 \int d^4x d^4y
K^\mu(x)  {\cal G}_{\mu \nu}(x-y)  K^{\nu}(y)} \nonumber\\
& \times & \exp{\frac i 2 \int d^4x d^4y J^\mu(x)  {\cal G}_{\mu \nu}(x-y)  
J^{\nu}(y)} \nonumber\\
 & \times & \exp{ i  \int d^4x d^4y
J^\mu(x)  \widetilde{\cal D}_{\mu \nu}(x-y)  K^{\nu}(y)} \nonumber\\ 
 &\times &\exp{\left(\ - i \int dx^{+} H_{cur}\right)},
\eey 
with the propagators
\bml
\bey
\fl {\cal G}_{\mu \nu} & = & \left( - g_{\mu \nu} + (n_\mu
\partial_{\nu} + n_\nu \partial_{\mu})(n \cdot
\partial)^{-1}\right) D_F 
- n_\mu n_\mu (n \cdot {\partial})^{-1} (n \cdot {\partial})^{-1},
\label{Gmunu}\\ 
\fl \widetilde{\cal D}_{\mu \nu} & = & \epsilon_{\mu \nu \lambda \rho} n^\lambda
\partial^\rho (n \cdot \partial)^{-1} D_F, \label{elmagpropag}
\eey
\eml 
where $n_\mu$ is the null vector $(n_{+} = 1, \bar{n} = 0)$, thus $(n
\cdot \partial)^{-1} = (\partial_{-})^{-1}$. The instantaneous part in (\ref{Gmunu}) can be removed if we choose the current Hamiltonian as 
\be \label{choiceofcurham}
H_{cur} = \frac 1 2 \int d^3 x \left[(\partial_{-})^{-1}
J^{+}\right]^2 + \frac 1 2 \int d^3 x \left[(\partial_{-})^{-1}
K^{+}\right]^2.
\ee
This leads to the effective propagators for the electric-electric and
magnetic-magnetic sectors 
\be \label{Dpropagator}
{\cal D}_{\mu \nu} = \left( - g_{\mu \nu} + (n_\mu
\partial_{\nu} + n_\nu \partial_{\mu})(n \cdot
\partial)^{-1}\right) D_F,
\ee
which have the form of the Abelian gauge field propagator for the LC 
gauge condition: $n_\mu  A^\mu = A^{+} = A_{-}$. 
Our propagators depend on
the null vector $n^\mu$, which enters into the theory due to the LF
quantization procedure. However, when the external currents are
conserved, then the dependence on $n^\mu$ disappears for the propagator
${\cal D}_{\mu \nu}$ but remains for the propagator $\widetilde{\cal D}_{\mu \nu}$. This means that the Lorentz invariance is broken by the
quantum interaction of electric and magnetic currents. This is a
real challenge to prove that for the physical observables (like cross
sections etc.) one may restore the Lorentz symmetry. Within the ET
formulation, one may prove \cite{11}, \cite{12}, \cite{13}, that the
dependence on a space like vector $n^\mu$ finally disappears. We hope
that a similar phenomenon happens also for (\ref{Dpropagator}), though
here we have a dependence on a null vector $n^2 = 0$. 

The Fourier transform of these propagators contains the Principle
Value (PV) prescription for the noncovariant pole $\dst PV \frac{1}{k_{-}} $. It is known, within the ET formulation, that this
prescription is not consistent for a nonAbelian gauge field theory
\cite{Bassetto1991}. 
In our gauge invariant procedure, we see that the PV prescription
is the only possibility for solving the LF constraint equations
in terms of a real valued distribution, at fixed $x^{+}$. Thus the
status of the PV prescription is consistent here within the LF
canonical formulation of Maxwell's theory.

The form of our perturbative propagators strongly suggests, that there
should be some gauge field model, which effectively produces the same
path-integral as we have found here. Since there are non local
interaction terms in the current Hamiltonian (\ref{choiceofcurham}),
thus we may add two auxiliary path integral variables $A_{+}$ and
$C_{+}$ into the phase space path-integral (\ref{magpathint1}). This
allows us to write the equivalent expression for the generating functional
\bml
\bey\label{Eq42a}
\fl Z[J^{+}, J^{i}, K^{+}, K^{i}] & = & \int {\cal D} A_i \ {\cal D}
A_{+} \ {\cal D} C_{+} \ \times \nonumber\\
&& \times \exp{\ i \int 
d^4x \left[ {\cal L}_{local} + C_{+} K^{+} + A_{+}J^{+} + A_i (J^i +
\epsilon_{ij}K^j) \right]},
\eey 
where the local Lagrangian density ${\cal L}_{local}$ is 
\be
\fl {\cal L}_{local} = \frac 1 2 \left( \partial_{-} A_{+}\right)^2 +
\left(\partial_{+} A_i - \partial_i A_{+}\right) \partial_{-} A_i 
+ \frac 1 2 \left( \partial_{-} C_{+}\right)^2 +
\epsilon_{ij} \partial_i A_j \partial_{-} C_{+}.\label{lagrLC}
\ee 
\eml 
In these expressions we have changed our gauge invariant path-integral
variables ${\cal A}_i$ into $A_i$, since here they are just dummy
variables of path integrations. 

In next step we treat this local Lagrangian density ${\cal L}_{local}$
as the case of the double LC gauge condition $A_{-} = C_{-} = 0$  imposed on the gauge 
invariant Lagrangian density
\bey \label{gaugeinvlagr}
{\cal L}_{inv} & = & \frac 1 2 \left( \partial_{+} A_{-} - \partial_{-}
A_{+}\right)^2 + \left(\partial_{+} A_i - \partial_i A_{+}\right)
\left( \partial_{-} A_i  -  \partial_{i} A_{-}\right) \nonumber\\
& - & \frac 1 2 \left(\epsilon_{ij} \partial_i A_j\right)^2
+ \frac 1 2 \left(\partial_{+} C_{-} - \partial_{-} C_{+}
- \epsilon_{ij} \partial_i A_j\right)^2,
\eey
where we have two independent Abelian gauge transformations:
\bml
\bey
A_\mu & \to & A_\mu + \partial_\mu \Theta_e, \label{eq45a}\\
C_{\pm} & \to & C_{\pm} + \partial_{\pm} \Theta_m. \label{eq45b}
\eey
\eml
The first three terms in (\ref{gaugeinvlagr}) form a standard LF
Abelian Lagrangian with the gauge potentials $A_\mu$, while last term
describes the contribution of the dual potentials $C_\pm$. We stress
that without this last term one cannot consistently quantize Maxwell's
theory with both electric and magnetic currents. 

Since in previous sections, we kept our canonical procedure explicitly
duality invariant, thus we would like to check whether (\ref{gaugeinvlagr})
has some duality symmetry. We find that, this Lagrangian density is invariant
(up to a total derivative) under the following transformation
\bml
\bey
{A}_i \longmapsto - \epsilon_{ij} {A}_j + (\partial_{-})^{-1}\left(\partial_i C_{-} + \epsilon_{ij} \partial_j A_{-}\right) ,\label{dualityAi}\\
A_{\pm} \longmapsto C_{\pm} \longmapsto - A_{\pm}.\label{dualityApm}
\eey
\eml
Since for the double LC gauge condition $A_{-} = C_{-} = 0$, these
transformations reduce to our former gauge invariant duality
transformation (\ref{gaugeinvduality}), thus we argue that they are
the generalized LF electromagnetic duality transformations for
the gauge and dual potentials. 
 
With the transformations (\ref{dualityAi}) and (\ref{dualityApm}),
supplemented with the transformation of currents (\ref{Eq4}), we
can check that the duality symmetry (up to a total derivative) is
possessed by the interaction Lagrangian density 
\bey 
{\cal L}_{int} & = & A_{-}\left( J^{-} - \epsilon_{ij} (\partial_{-})^{-1} \partial_{i} K^{j} \right) + C_{-}\left( K^{-} + (\partial_{-})^{-1} \partial_{i} K^{i} \right) \nonumber\\
&+ &A_{+} J^{+} + C_{+} K^{+} + A_i\left(J^{i} + \epsilon_{ij} K^{j}\right).\label{intLagr}
\eey
Apparently this interaction Lagrangian is not satisfactory, since
the gauge and dual potentials couple nonlocally to external currents. 
We could introduce further auxiliary variables to keep all
potential-currents couplings local but at the price of introducing the
Lagrange multiplier fields - this would be another manifestation of
the canonical constraint (\ref{canconstr}). However we have decided to
not proceed in this direction here and to discuss it elsewhere.

Though the interaction Lagrangian (\ref{intLagr}) looks strange, 
it behaves properly under the gauge transformations. From the
transformation (\ref{eq45a}) one gets the electric current
conservations law $\partial_\mu J^{\mu} = 0$, while from   
(\ref{eq45b}) one gets the magnetic current conservation law
$\partial_\mu K^{\mu} = 0$. This convinces us that
we may treat the gauge invariant Lagrangian (\ref{gaugeinvlagr})
with the interaction Lagrangian (\ref{intLagr}) as a good starting point
for the canonical quantization procedure with different
gauge conditions \cite{Dzimida04}.

\section{Conclusions and further prospects}

In this paper, we show that the electromagnetic duality can be 
consistently defined for the LF Maxwell theory with the
classical external electric and magnetic currents. For practical
reasons we introduce the independent physical fields ${\cal
A}_{i}$ which allow to satisfy identically the LF electric and magnetic
Gauss law equations. Remembering that fields ${\cal A}_{i}$ 
are not the transverse components of some gauge potential, 
we are satisfied that for these physical modes the LF duality
transformation looks like the Susskind conjecture ${\cal A}_i \to - \epsilon_{ij} {\cal A}_j$.

We  propose a novel canonical LF procedure, which is based on the
longitudinal translation operator $P^{+}$, which, being a kinematic
generator, has the same form both for free and interacting theories.
Usually the canonical procedure starts with a Lagrangian, but in the
case of Maxwell's theory with electric and magnetic currents, the
starting point is Maxwell's equations. In our novel canonical
procedure, we can safely take $P^{+}$ from the free field theory,
while the LF Hamiltonian $P^{-}$ can be found as a solution of the
functional differential equation. 

We also use the path-integral formulation for achieving two
different goals. First, we find the perturbative
propagators with the instantaneous (in $x^{+}$) terms, which then can 
be cancelled by the contribution from the LF Hamiltonian. The
effective propagators (\ref{elmagpropag}, \ref{Dpropagator}) 
have the form known from Schwinger's source theory
\cite{Schwigersource} for an arbitrary  constant vector $n^\mu$.
Here we present their LF canonical derivation for the case of a
null vector $n^2 = 0$. 
Second, we prove that our gauge invariant formalism is equivalent
to some gauge theory with two potentials. We show that the LC gauge
condition can be chosen for both the gauge and dual potentials. Our  
two gauge potential Lagrangians (\ref{gaugeinvlagr}) and (\ref{intLagr})
are far simpler than the one proposed long ago by  Zwanziger
\cite{Zwanziger1971} and recently used for the LF quantization by
Mukherjee and Bhattacharya \cite{Asmita2000}. Here we would like to
point out that our result disagrees with \cite{Asmita2000}, where the
Hamiltonian contains an additional instantaneous term (for the
interaction of electric and magnetic currents), which explicitly
breaks the rotational symmetry. In our case the rotational symmetry is
broken solely by a choice of the LF surface.

The two gauge potential Lagrangian densities (\ref{gaugeinvlagr}) and
(\ref{intLagr}) can form a starting point for the dual formulation
of Abelian theory within the LF approach. When further such
formulation is generalized to a nonAbelian theory, then a very
encouraging possibility arises - the LF version of the dual
superconductor models of color confinement (for introduction and
earlier references see \cite{Ripka2004}). 

When one performs the perturbative calculations, with currents 
generated by some charged matter fields, then inevitably the
perturbative propagators (\ref{elmagpropag}, \ref{Dpropagator}) will
produce momentum integrals with the UV divergences. Therefore one will
have to regularize the model without spoiling the gauge and duality
invariance. Since the antisymmetrical symbol $\epsilon_{ij}$ appears
explicitly in many expressions, then the dimensional regularization
seems to be impractical here and one should look for a kind of the
Pauli-Villars regularization. Possibly, this can give rise to the
regularized Lagrangian density with higher derivative terms (see \cite{Panagiotakoulos}). 

When the charged matter is treated as a part of the quantum dynamical
system, then one may check the second Susskind conjecture
\cite{Susskind96}, that in the LF approach there is no need for any
other non localities, like Dirac strings, beside the usual
$(\partial_{-})^{-1}$ integral operator. Such investigations should
also answer another question how the Dirac-Schwinger quantization
condition for electric and magnetic charges \cite{Dirac1931},
\cite{DiracSchwinger2} appears within the LF formalism. We hope to give definite answers to these problems in our future publication.

\appendix

\section{The LF notation}

We use the natural units $c = \hbar = 1$. Our LF notation starts
 with the definitions of null components for
the coordinates $x^{\pm} = (x^{0} \pm x^1)/\sqrt{2}$,
while the transverse components are $x^i = (x^2, x^3)$. The similar
definitions are taken for any 4-vectors. The LF surface coordinates
are denoted as $\bar{x} = (x^{-}, x^i)$. 
The partial derivatives
are taken with respect to contravariant coordinates, thus we have
$\partial_{+} = \partial/\partial x^{+}, \partial_{-} =
\partial/\partial x^{-}, \partial_{i} = \partial/\partial x^{i}$. The metric
tensor has non vanishing components $g_{+-} = g_{-+} =1, g_{ij} =
- \delta_{ij}$. The scalar product of 4-vectors is $a \cdot b =
a_{+} b_{-} + a_{-} b_{-} - a_i b_i$, while for the LF surface components 
we have ${\bar a} \cdot {\bar b} = a^{-} b_{-} - a_i b_i$.
There are two natural antisymmetric tensors
$\epsilon_{ij} = - \epsilon_{ji}$ with $\epsilon_{23} = 1$, and
$\epsilon_{+-ij} = \epsilon_{ij}$.

The inverse differential operator $(\partial_{-})^{-1}$ is taken as
the distribution
\be
(\partial_{-})^{-1}(\bar{x}) = \frac 1 2 {\rm sgn}(x^{-}) \delta^2(x^i),
\ee  
which means that we use the PV prescription in its Fourier integral representation
\be
(\partial_{-})^{-1}(\bar{x}) = - i \int \frac{d^3{\bar k}}{(2 \pi)^3}
e^{ i \bar{x} \cdot \bar{k}} {\rm PV} \frac{1}{k_{-}}. 
\ee
Another integral operator $\partial_i \Delta^{-1}_\perp$ is given as
\be
\partial_i \Delta^{-1}_\perp(\bar{x}) = \frac{1}{2\pi} \frac{x^i}{x_\perp^2} \delta(x^{-}), \qquad x_\perp^2 = x^i x^i, 
\ee
while the covariant propagator function is defined as
\be
D_{F}(x) = \int d^4k \frac{e^{- ik \cdot
x}}{2k_{+} k_{-} - k_\perp^2 + i
\epsilon}. 
\ee


\section{Dirac's strings within the ET and LF formulations}\label{appstrings}

Dirac's approach with strings \cite{Dirac1931}, \cite{Dirac1948} is
basically a gauge potential formulation, where the modifications in
Bianchi identities (2b) are compensated by the redefinition of the
electromagnetic field tensor
\be \label{Diracstringmod}
F_{\mu \nu} = \partial_\mu A_{\nu} - \partial_\nu A_{\mu} + \frac 1 2 \epsilon_{\mu \nu \lambda \rho} G^{\lambda \rho},
\ee
provided the string tensor $G^{\lambda \rho}$ satisfies the equation
\be
\partial_\lambda G^{\lambda \rho} = K^\rho. \label{defG}
\ee
A straight line string is given by 
\be
G^{\lambda \rho} = (n \cdot \partial)^{-1} \left( n^\lambda K^\rho
- n^\rho K^\lambda\right)
\ee
where $n^\mu$ is a given fixed vector. In the ET approach one chooses
this fixed vector to be space-like $n^\mu = (0, \vec{n})$, thus the
integral operator $ (n \cdot \partial)^{-1} = - \left(\vec{n} \cdot\vec{\nabla}\right)^{-1}$ does not include temporal evolution.
The orientation of $\vec{n}$ is arbitrary and no physical result
should depend on it.  

This procedure can be also applied in the LF approach and one finds
that there are 2 different possibilities - one can take either $\left(n^{-} = 1, n^{+} = n^i = 0\right) $ or $\left(n^{+} = n^{-} =0, n^i \neq 0\right)$ - which lead to no temporal evolution in
the integral operator $ (n \cdot \partial)^{-1}$. In the first case
one has a fixed null vector $n^2 = 0$, while in the second case one
again has a space-like vector $n^2 < 0$. 

However we stress that equation (\ref{defG}) is not duality invariant since
it contains only the magnetic sources $K^\rho$. Since our present
paper is devoted to the gauge and duality invariant formulation we
will not discuss such Dirac strings any longer here but postpone a
more detailed analysis to a future publication.

Within the Hamiltonian formulation, one can adopt Dirac's idea
(\ref{Diracstringmod}) for transforming  inhomogeneous constraint
equations into homogeneous equations. One may define the electromagnetic field tensor as  
\be
F_{\mu \nu} = {\cal F}_{\mu \nu} +  n_\mu f_{\nu} - n_\nu f_{\mu} +  \epsilon_{\mu \nu \lambda \rho} n^{\lambda} g^{\rho},
\ee
where the components of ${\cal F}_{\mu \nu}$ satisfy the homogeneous
constraint equations, while $f^\mu$ and $g^\mu$ are some functions
which may depend on electric and magnetic charges. This modification
is duality invariant, provided one introduces the duality transformation for $f^\mu$ and $g^\mu$
\be
g^\mu \mapsto f^\mu \mapsto - g^\mu.
\ee
The fixed vector $n^\mu$ is quite arbitrary and here we will discuss
only the simplest choices. In the ET approach one can take $\left(n^0 = 1, \vec{n} = 0\right)$ which leads to
\bey 
\vec{E} &  = & \vec{\cal E} + \vec{f},\qquad \vec{\nabla} \cdot \vec{f} = J^0,
\label{ETstrings1}\\
\vec{B} &  = & \vec{\cal B} - \vec{g},\qquad \vec{\nabla} \cdot \vec{g} = - K^0.
\label{ETstrings2}
\eey
In the LF approach one can take $\left(n^{-} = 1, n^{+} = n^i = 0\right)$ which leads to
\bey 
E_{-} &  = & {\cal E}_{-} - f_{+},\qquad \partial_{-}{f}_{+} = J^{+},\label{LFstrings1}\\
B_{-} &  = & {\cal B}_{-} + g_{+},\qquad \partial_{-}{g}_{+} = - K^{+},\label{LFstrings2}\\
E_{i} &  = & {\cal E}_{i}, \qquad B_{i}   =  {\cal B}_{i}.\label{LFstrings3}
\eey  
These two choices of a constant vector $n^\mu$ are equivalent, since
they indicate the respective temporal evolution parameters:  in the ET
case we have $n_\mu x^{\mu} = x^0$, while in the LF case we 
have  $n_\mu x^{\mu} = x^{+}$. Evidently equations
(\ref{LFstrings1}-\ref{LFstrings3}) are the solutions
(\ref{eq:28a}-\ref{eq:28c}) proposed in the main text, while 
equations (\ref{ETstrings1}-\ref{ETstrings2}) lead to a Coulomb-like solution
for electric and magnetic monopoles.


\end{document}